\documentclass[conference]{IEEEtran}
\IEEEoverridecommandlockouts
\usepackage{cite}
\usepackage{amsmath,amsthm,amssymb,amsfonts,dsfont}
\usepackage{algorithmic}
\usepackage{graphicx}
\usepackage{textcomp}
\usepackage{xcolor}
\usepackage{soul}
\usepackage{tikz}
\usepackage{graphics}
\usepackage{pgfplots}
\usepackage{mathtools, cuted} %to span big equations
\usepackage[ruled,vlined]{algorithm2e}
\usepackage{xcolor}

\usepgfplotslibrary{groupplots}
\usetikzlibrary{backgrounds,calc,shadings,shapes,arrows,arrows.meta,shapes.arrows,shapes.symbols,patterns, datavisualization, datavisualization.formats.functions, matrix}
\def\BibTeX{{\rm B\kern-.05em{\sc i\kern-.025em b}\kern-.08em
    T\kern-.1667em\lower.7ex\hbox{E}\kern-.125emX}}

{\catcode`,=\active
  \gdef,{\mathcomma\discretionary{}{}{}}
}

\allowdisplaybreaks %in order to split big formulas across different breaks, e.g., pages, or columns
\pgfplotsset{compat=1.14}
\usepackage{url}

\begin{document}

\title{WIP: Short-Term Flow-Based Bandwidth Forecasting using Machine Learning
}

\author{\IEEEauthorblockN{Maxime Labonne, Jorge L\'opez, Claude Poletti, Jean-Baptiste Munier
\IEEEauthorblockA{%\textit{dept. name of organization (of Aff.)} \\
\textit{Airbus Defence and Space}\\
Issy-Les-Moulineaux, France \\
\{maxime.labonne, jorge.lopez-c, claude.poletti, jean-baptiste.munier\}@airbus.com
}}}
\maketitle

%%%%%%%%%%%%%%%%%%%%%%%%%%%%%%%%%%%%%%%%%%%%%%%%%%%%%%%%%%%%%%%%%%%%%%%%%%%%%%%%%%%%%%%%%%%%%%%%%%%%%%%%%%

\begin{abstract}
This paper proposes a novel framework to predict traffic flows' bandwidth ahead of time. Modern network management systems share a common issue: the network situation evolves between the moment the decision is made and the moment when actions (countermeasures) are applied. This framework converts packets from real-life traffic into flows containing relevant features. Machine learning models, including Decision Tree, Random Forest, XGBoost, and Deep Neural Network, are trained on these data to predict the bandwidth at the next time instance for every flow. Predictions can be fed to the management system instead of current flows bandwidth in order to take decisions on a more accurate network state. Experiments were performed on 981,774 flows and 15 different time windows (from 0.03s to 4s). They show that the Random Forest is the best performing and most reliable model, with a predictive performance consistently better than relying on the current bandwidth (+19.73\% in mean absolute error and +18.00\% in root mean square error). Experimental results indicate that this framework can help network management systems to take more informed decisions using a predicted network state.
\end{abstract}


\newtheorem{proposition}{Proposition}
\newtheorem{definition}{Definition}

%%%%%%%%%%%%%%%%%%%%%%%%%%%%%%%%%%%%%%%%%%%%%%%%%%%%%%%%%%%%%%%%%%%%%%%%%%%%%%%%%%%%%%%%%%%%%%%%%%%%%%%%%%

\section{Introduction}\label{sec:intro}

Recent approaches to network management, such as Software Defined Networking~(SDN), enable dynamic and flexible network control to improve performance and monitoring. This paradigm can take decisions from observations from the network and reconfigure it for performance, troubleshooting, or security issues. However, there is a delay between the moment a decision is made and the moment it is applied. It means that the reconfiguration might be suboptimal, since it was chosen on past (and maybe outdated) information.

A solution to this problem is to take decisions on the future network state, when the decision is applied. This state can be forecasted knowing the current state and how it typically evolves. Machine learning models can be trained to learn this behavior and predict the next state. In this work, we focus on predicting the future bandwidth of every flow in the network. As a practical and concrete example, this prediction may then be used with the goal of taking admission and routing decisions based on the flows' priorities, as proposed in \cite{lopez2020priority}.

Achieving this goal requires two conditions for bandwidth forecasting: it must be \emph{short-term} and \emph{flow-based}. Short-term forecasting (from milliseconds to minutes), in contrast to long-term forecasting (from hours to years), cannot rely on seasonality and requires more data. Flow-based bandwidth is extremely user- and application-dependent with an important stochastic component.

In this paper, we evaluate 4 different machine learning algorithms: Decision Tree~(DT), Random Forest~(RF), XGBoost~(eXtreme Gradient Boosting), and Deep Neural Network~(DNN). A real-life traffic dataset (CAIDA Anonymized Internet Traces 2016 Dataset \cite{caida}) is converted into flows with specific, relevant features. Machine learning models are trained on this new dataset to predict the required bandwidth of every flow at the next time instance. Extensive feature engineering and parameters optimization lead to predictions that are significantly better than taking decisions on the current network state.

The remainder of the paper is organized as follows. Section~\ref{sec:related} discusses related work in the area of short-term bandwidth prediction. Section~\ref{sec:preprocessing} describes the preprocessing stage, from the feature selection to the creation of the flow dataset. Section~\ref{sec:results} presents the machine learning and the experimental results on this dataset. Finally, Section~\ref{sec:conclusion} discusses areas of future work and concludes this paper.

%%%%%%%%%%%%%%%%%%%%%%%%%%%%%%%%%%%%%%%%%%%%%%%%%%%%%%%%%%%%%%%%%%%%%%%%%%%%%%%%%%%%%%%%%%%%%%%%%%%%%%%%%%

\section{Related Work}\label{sec:related}
The topic of network traffic prediction is popular in the literature for its numerous applications. Long-term predictions are mainly used to forecast capacity requirements while short-term predictions are used for dynamic resource allocation. The latter category is often modelled as a binary classification problem, where the goal is to categorize flows as elephants (very large flows) or mice.

Jahnke et al. (2018) \cite{jahnke} claim to be the first approach to fine-grained per-flow traffic prediction. The authors use a Frequency-based Kernal Kalman Filter~(FKKF) changing the operating space from time to frequency space. They compare this approach with 
Autoregressive Integrated Moving Average~(ARIMA) and Generalized Autoregressive Conditional Heteroskedasticity~(GARCH) models on a real-life traffic dataset with 20 flow groups. This approach achieves 10.9\% prediction error on average on an optimal interval of 0.49 seconds (ARIMA = 77.3\% and GARCH = 95.2\%).

Hardegen et al. (2019) \cite{hardegen} present a framework for flow-based throughput classification using deep neural networks. They provide a real-life dataset of 252 million flows collected during one week and a comprehensive analysis using t-distributed Stochastic Neighbor Embedding~(t-SNE). Their goal is to classify the predicted bitrates into three classes instead of elephant and mice flows. Three hyperparameters are optimized: number of nodes, number of layers, and learning rate. Experiments show that this forecasting achieves an average accuracy of 82\% within a continuous interval of one week.

Lazaris and Prasanna (2019) \cite{lazarisb} analyse the performance of Long short-term memory~(LSTM) networks and ARIMA models to predict link throughputs on a real-life traffic dataset (CAIDA Anonymized Internet Traces 2016 Dataset). They evaluate three variations of each model on four epoch durations (5 sec, 10 sec, 15 sec, 30 sec) and a 50-50 train/test split. LSTM networks obtain a mean average error significantly lower than ARIMA models in every scenario, especially the vanilla LSTM.

Our solution benchmarks various machine learning algorithms, including tree-based models that are absent in previous work. Differently from existing works, our goal is to predict the exact future bitrate of every flow (regression, not classification), including 0 bit/s (stopped flows). Finally, the prediction can be extremely short with a time window of just 0.03 sec.

%%%%%%%%%%%%%%%%%%%%%%%%%%%%%%%%%%%%%%%%%%%%%%%%%%%%%%%%%%%%%%%%%%%%%%%%%%%%%%%%%%%%%%%%%%%%%%%%%%%%%%%%%%

\section{Preprocessing}\label{sec:preprocessing}

\subsection{Dataset}

The CAIDA Anonymized Internet Traces 2016 dataset \cite{caida} was chosen for several reasons: i) it contains a large quantity of data to ensure correct learning even for deep neural networks; ii) it represents real-life traffic; iii) it is recent and contains few errors and iv) it was already studied in the literature \cite{lazarisb}. Data were collected from high-speed monitors on a commercial backbone link and then anonymized. It provides raw Internet Protocol~(IP) packet captures (PCAP files) that can be processed to extract the most relevant features for this problem.

However, the extreme diversity of traffic in this dataset (Tier-1 ISP) makes prediction more difficult than on a type of traffic with few applications. Its high throughput also means that one second of traffic represents a larger number of packets (2.6 million on average) compared to a traditional network. 

\subsection{Feature extraction from raw packets}

The feature extraction process is performed using TShark or Wireshark. Wireshark is a well-known network protocol analyzer that is used across many commercial and non-profit projects from enterprises, government agencies, and educational institutions. The same packet format is used to convert packets into flows in the remainder of the paper. Formally, we consider a flow as:

\begin{definition}\label{def:flow} A traffic flow or flow is a series of IP packets during a certain time interval, which share five common features: \emph{source IP address}, \emph{destination IP address}, \emph{source port}, \emph{destination port}, and \emph{protocol}.\end{definition}

IP packets contain a lot of information correlated to bandwidth. A number of fields have been identified to contribute to the overall prediction:
\begin{itemize}
\item $Source/destination$ $IP$ $address,$ $source/destination$ $port$ $number,$ $protocol$ -- 5-tuple used to identify packets that belong to the same flow;
\item $Time$ -- seconds elapsed since start of the capture;
\item $Delta time$ -- seconds between this packet and the previous packet;
\item $DSCP$ -- (Differentiated Services Code Point) packet classification to provide Quality of Service~(QoS);
\item $Length$ -- packet size in bytes, necessary to calculate the bandwidth;
\item $Flags$ -- TCP flags (NS, CWR, URG, ACK, PSH, RST, SYN, FIN, ECN) can be helpful to understand the behavior of the connection (frequent resets, numerous urgent packets, etc.).
\end{itemize}

Other fields can be used to anticipate a state of congestion and thus to refine the prediction of the machine learning algorithm in cases close to congestion or in congestion. The purpose of this process is not to determine or predict congestion, but to enrich the prediction analysis with new features:

\begin{itemize}
\item $TCPWindowSize$ -- size of the receive window (not the congestion window), which is the amount of data that a computer can accept without acknowledging the sender;
\item $TCPWindowScale$ -- option to increase the receive window size allowed in TCP above its former maximum value of 65,535 bytes (used for efficient data transfer in long fat networks);
\item $TCPRetransmission$ -- a stream is retransmitted if it is not acknowledged (true if this field is not zero).
\end{itemize}

The congestion window cannot be extracted since it is computed on the operating system. It could be estimated, but its implementation is system-dependent which is why it is not considered in this paper.

The dataset after this feature extraction process contains 10 seconds of traffic, comprised of 981,774 flows and 26,074,447 packets.

\subsection{Feature engineering}

The goal of this process is to aggregate packets with the previous format into traffic flows with the most important features for predicting the bandwidth evolution. This problem could be treated as a time series forecasting problem with a lookback window of $n$ previous flow observations. However, we argue that i) the frequent appearance of new flows creates a sparse dataset (58\% of flows are new flows with a lookback of just 1s) and ii) most previous flows features are not relevant to predict the next time slot's bitrate.

Time series-related features are instead embedded to provide information about the traffic's history: total number of traffic flows, flow's cumulative bitrate, past bitrate and total number of packets. A time window (between 0.03s and 4s) is set to split the 10 seconds of traffic into slots of $x$ seconds. Mathematical functions such as max, min, average and standard deviation (std, for short) are used on packet features to create additional flows features. Table \ref{tab:features} shows the comprehensive list of 80 flows features.

\begin{table}[!htb]
    \centering
    \caption{Description of flows features.}
    \begin{tabular}{l|l}
\hline
\textbf{Feature}  & \textbf{Description}                            \\ \hline
ID                & SrcIP, DstIP, SrcPort, DstPort, and protocol    \\ \hline
FlowCount         & Number of flows created so far                  \\ \hline
Protocol          & IPv4 payload protocol (e.g., TCP, UDP, ICMP)    \\ \hline
DSCP              & Common packet DSCP                              \\ \hline
Timeslot          & Number of time windows elapsed                  \\ \hline
PacketTotal       & Number of packets in this flow                  \\ \hline
Length            & Cumsum, max, min, average, std packet length    \\ \hline
Flags\footnotemark
                  & Total, max, min, average, std packet flags      \\ \hline
Deltatime         & Max, min, average, std deltatime                \\ \hline
TCPWindowSize     & Max, min, average, std TCPWindowSize            \\ \hline
TCPWindowScale    & Max, min, average, std TCPWindowScale           \\ \hline
TCPRetransmission & Total, max, min, average, std TCPretransmission \\ \hline
CumBitrate        & Max, min, sum, average cumulative bitrate       \\ \hline
Bitrate           & Current bitrate and past bitrate                \\ \hline
BitratePast       & Last timeslot's bitrate                         \\ \hline
BitrateFuture     & Next timeslot's bitrate                         \\ \hline
\end{tabular}
\label{tab:features}
\end{table}
\footnotetext{Performed for each packet flag: ECN, NS, CWR, URG, ACK, PSH, RST, SYN, FIN.}

%%%%%%%%%%%%%%%%%%%%%%%%%%%%%%%%%%%%%%%%%%%%%%%%%%%%%%%%%%%%%%%%%%%%%%%%%%%%%%%%%%%%%%%%%%%%%%%%%%%%%%%%%%

\section{Experimental results}\label{sec:results}

\subsection{Metrics}

Two metrics are used to evaluate the performance of each model: mean absolute error and root mean square error.

\textbf{Mean Absolute Error}~(MAE) measures the average error in a set of predictions, without considering their direction. It is expressed in units of the variable of interest. This measure is defined as follows:

\begin{equation}
MAE = \frac{1}{n}\sum_{j=1}^{n}\left | y_{j} - \hat{y}_{j} \right |
\end{equation}

where $n$ is the number of samples and $\hat{y}_{j}$ and $y_{j}$ are the predicted and the real values, respectively.

\textbf{Root Mean Square Error}~(RMSE) is a quadratic scoring rule that gives a high importance to large errors. It is also expressed in units of the variable of interest. This measure is defined as follows:

\begin{equation}
RMSE = \sqrt{\frac{1}{n}\sum_{j=1}^{n}\left ( y_{i} - \hat{y}_{i} \right )^{2}}
\end{equation}

%%%%%%%%%%%%%%%%%%%%%%%%%%%%%%%%%%%%%%%%%%%%%
\subsection{Experiments}

Four machine learning algorithms are selected for this regression task. They are comprised of three tree-based models (DT, RF, XGBoost), and one DNN. All parameters presented in this subsection have been optimized manually or through random search. These models are trained on a Ryzen 5 1600 CPU and a GTX 1080 GPU with 16GB of RAM. Categorical features are one-hot encoded and a random 80/20 train-test split is used.

\textbf{Decision Tree} is a supervised learning technique used for classification and prediction. Decision Trees are structured as binary trees, where each node represents a test on a feature and each leaf node holds an output. The Decision Tree algorithm used in this work is CART~(Classification and Regression Trees) with a maximum depth of 12. It was trained in 1min 5s (37µs/sample) on average for a time window of 1.0s.

\textbf{Random Forest} uses numerous relatively uncorrelated decision trees operating as an ensemble. The assumption is that this ensemble model outperforms any of the individual constituent trees. Indeed, the trees are uncorrelated, which means they protect each other from their individual errors. Our implementation uses 30 individual trees with a maximum depth of 10 to create the forest. It was trained in 6min 47s (232µs/sample) on average for a time window of 1.0s.

\textbf{XGBoost} \cite{Chen_2016} combines the results of a set of simpler and weaker tree models in order to provide a better prediction. Unlike the RF, this algorithm works sequentially. 100 gradient boosting trees are used with a maximum depth of 20, a learning rate of 0.01, an $alpha$ (L1 regularization term on weights) of 1, and a $colsample\_bytree$ (subsample ratio of columns when constructing each tree) of 0.9. This model was trained in 11min 1s (378µs/sample) on average for a time window of 1.0s.

\textbf{Deep Neural Network} is a feedforward multilayered neural network. Our implementation has 4 hidden layers (256-128-64-32) and uses ReLu as the activation function. It also has a cyclical learning rate (between 0.00001 and 0.001 with a step size of 1000000) to gradually modify its value. It is trained on 10 epochs with a batch size of 64 in 19min 6s (665µs/sample) on average for a time window of 1.0s.

Two other boosting techniques have been tested: CatBoost and LightGBM. Their performance is inferior to that of XGBoost, which is why they have been excluded from this section. Likewise, a modified Transformer architecture without positional encoding has been studied. However, this model does not learn the traffic's behavior and converges around the same average bitrate for every flow.

% Likewise, two Transformer architectures have been studied: a vanilla Transformer and TabNet \cite{arik2020tabnet}. The former converged around an average bitrate for every flow, while the latter never obtained credible results.

\subsection{Results}

The given results are an average of 5 repetitions (training and testing) for each model and each time window. The inference (prediction) times for a time window of 1s are 0.5s (1µs/sample), 0.8s (2µs/sample), 4.3s (10µs/sample) and 8.1s (19µs/sample) on average for the DT, RF, XGBoost and DNN models, respectively.

Table \ref{tab:mae} and Table \ref{tab:rmse} show that the Random Forest outperforms the other models for almost every time window in both MAE and RMSE. XGBoost is the second best performing model, especially in RMSE compared to the Decision Tree. Finally, the DNN provides good results in RMSE but performs poorly in terms of MAE. These results have to be compared to the current bitrate ("Base" in the tables), which would be the bitrate given to the network management system without our forecasting solution.

\begin{table}[!htb]
    \centering
    \caption{MAE for different time windows and algorithms.}
    \begin{tabular}{l|l|l|l|l||l}
\hline
\multicolumn{1}{c|}{\textbf{Window}} & \textbf{DT} & \textbf{RF} & \textbf{XGB} & \textbf{DNN} & \textbf{Base} \\ \hline
\textbf{0.03s}                       & \textbf{24151.6}     & 25292.2     & 26890.8      & 33232.4      & 40812.1       \\ \hline
\textbf{0.05s}                       & 19168.1     & \textbf{17384}       & 19656.8      & 19630.5      & 23802.1       \\ \hline
\textbf{0.1s}                        & 11191.2     & \textbf{10856.2}     & 12539.9      & 11909.7      & 13872.6       \\ \hline
\textbf{0.2s}                        & 7835.97     & \textbf{7195.66}     & 7376.64      & 9092.48      & 8326.44       \\ \hline
\textbf{0.3s}                        & 6091.32     & \textbf{5634.69}     & 5858.82      & 6868.63      & 6414.17       \\ \hline
\textbf{0.4s}                        & 5136.11     & \textbf{4800.6}      & 5082.24      & 5943.07      & 5490.37       \\ \hline
\textbf{0.5s}                        & 5122.48     & \textbf{4460.52}     & 4849.74      & 6986.8       & 4973.05       \\ \hline
\textbf{0.6s}                        & 4951.06     & \textbf{4545.74}     & 4673.68      & 5541.52      & 4958.65       \\ \hline
\textbf{0.7s}                        & 4296.34     & \textbf{4039.78}     & 4402.57      & 5269.07      & 4569.34       \\ \hline
\textbf{0.8s}                        & 4195.43     & \textbf{3775.96}     & 3987.13      & 6504.16      & 4175.09       \\ \hline
\textbf{0.9s}                        & 3660.36     & \textbf{3548.98}     & 3686.54      & 4724.77      & 3985.11       \\ \hline
\textbf{1.0s}                          & 3434.06     & \textbf{3266.43}     & 3439.61      & 5238.23      & 3831.22       \\ \hline
\textbf{2.0s}                          & 2205.06     & \textbf{2060.66}     & 2259.93      & 4214.46      & 2856.29       \\ \hline
\textbf{3.0s}                          & 1532.15     & \textbf{1361.12}     & 1413.5       & 2013.34      & 2072.49       \\ \hline
\textbf{4.0s}                          & 1048.75     & \textbf{991.553}     & 1005.72      & 1808.74      & 1748.97       \\ \hline
\end{tabular}
\label{tab:mae}
\end{table}

\begin{table}[!htb]
    \centering
    \caption{RMSE for different time windows and algorithms.}
    \begin{tabular}{l|p{0.055\textwidth}|p{0.055\textwidth}|p{0.055\textwidth}|p{0.055\textwidth}||p{0.05\textwidth}}
\hline
\multicolumn{1}{p{0.05\textwidth}|}{\textbf{Window}} & \textbf{DT} & \textbf{RF} & \textbf{XGB} & \textbf{DNN} & \textbf{Base} \\ \hline
\textbf{0.03s}                       & 2.38E+05    & \textbf{2.27E+05}    & 2.63E+05     & 2.52E+05     & 3.35E+05      \\ \hline
\textbf{0.05s}                       & 2.32E+05    & \textbf{1.48E+05}    & 2.06E+05     & 1.52E+05     & 1.87E+05      \\ \hline
\textbf{0.1s}                        & 1.27E+05    & \textbf{1.06E+05}    & 1.49E+05     & 1.09E+05     & 1.23E+05      \\ \hline
\textbf{0.2s}                        & 1.64E+05    & 83844.7     & \textbf{83032.5}      & 84911.1      & 96803.7       \\ \hline
\textbf{0.3s}                        & 1.10E+05    & \textbf{70986.3}     & 71347.9      & 74344        & 83184.8       \\ \hline
\textbf{0.4s}                        & 77577.5     &\textbf{60315.8}     & 63442.7      & 66249.6      & 75468.3       \\ \hline
\textbf{0.5s}                        & 1.02E+05    & \textbf{61251.7}     & 71185.9      & 65714        & 72386.2       \\ \hline
\textbf{0.6s}                        & 91798.3     & 73486.3     & 75330.7      & \textbf{70782.7}      & 76646.1       \\ \hline
\textbf{0.7s}                        & 82699.1     & \textbf{62968.8}     & 74009.7      & 79267.2      & 74415.0         \\ \hline
\textbf{0.8s}                        & 1.01E+05    & 61712.3     & 73236.1      & \textbf{61054.8}      & 67031.3       \\ \hline
\textbf{0.9s}                        & 75434.3     & \textbf{60039.4}     & 67875.8      & 60788.5      & 66113.0         \\ \hline
\textbf{1.0s}                        & 64100.8     & \textbf{56669.9}     & 65588.7      & 62019.9      & 65335.4       \\ \hline
\textbf{2.0s}                        & 51425.1     & \textbf{40207.7}     & 42216.8      & 47909.9      & 57422.9       \\ \hline
\textbf{3.0s}                        & 60098.2     & 32575.5     & \textbf{32573.5}      & 60686        & 42297.7       \\ \hline
\textbf{4.0s}                        & 44988.6     & \textbf{24431.4}     & 22440.8      & 48345.7      & 38599.8       \\ \hline
\end{tabular}
\label{tab:rmse}
\end{table}

RF's performance can be further analyzed with relative MAE and RMSE, as defined by equation \ref{eq:relative}.

\begin{equation}\label{eq:relative}
Relative\:Error = 1-\frac{Error_{RF}}{Error_{Base}}
\end{equation}

An interesting observation can be made from  Table \ref{tab:engineering}: the average accuracy gain (on each time window) for the two best performing models, evaluated on a minimal set a features ($SrcIP$, $DstIP$, $SrcPort$, $DstPort$, $Protocol$, $DSCP$, $Bitrate$, $BitrateFuture$) and on the full set of features. Results show that the feature engineering process is essential to obtain useful predictions.

\begin{table}[!htb]
    \centering
    \caption{Average accuracy gain with and without feature engineering.}
    \begin{tabular}{l|c|c|c|c}
\cline{2-5}
                       & \multicolumn{2}{c|}{\textbf{Random Forest}}                                                     & \multicolumn{2}{c}{\textbf{XGBoost}}                                               \\ \cline{2-5} 
                       & \multicolumn{1}{c|}{\textbf{8 features}} & \multicolumn{1}{c|}{\textbf{80 features}} & \multicolumn{1}{c|}{\textbf{8 features}} & \multicolumn{1}{c}{\textbf{80 features}} \\ \hline
\textbf{Relative MAE}  & -0.23\%                                  & +19.73\%                                  & -6.15\%                                  & +14.53\%                                 \\ \hline
\textbf{Relative RMSE} & -21.16\%                                 & +18.00\%                                  & -41.52\%                                 & +7.82\%                                  \\ \hline
\end{tabular}
\label{tab:engineering}
\end{table}

\begin{figure}[!htb]
    \centering
    \begin{tikzpicture}[scale=1]
    \begin{axis}[
        xlabel=Time window (s),
        ylabel=Accuracy gain (\%),
        legend style={anchor=north east}
        ]

    \addplot[smooth,mark=x, red] plot coordinates {
        (0.03,38.0276928)
        (0.05,26.9644275)
        (0.1,21.7435809)
        (0.2,13.5805939)
        (0.3,12.1524687)
        (0.4,12.5632699)
        (0.5,10.3061501)
        (0.6,08.3270648)
        (0.7,11.5894199)
        (0.8,09.5597939)
        (0.9,10.943989)
        (1.0,14.7417794)
        (2.0,27.8553648)
        (3.0,34.3244117)
        (4.0,43.3064604)
    };
    \addlegendentry{Relative MAE}
    
    \addplot[smooth,mark=+, blue] plot coordinates {
        (0.03,32.238806)
        (0.05,20.855615)
        (0.1,13.8211382)
        (0.2,13.386885)
        (0.3,14.6643377)
        (0.4,20.0779665)
        (0.5,15.3820756)
        (0.6,04.1225842)
        (0.7,15.3815763)
        (0.8,07.935099)
        (0.9,09.1866955)
        (1.0,13.2631009)
        (2.0,29.9796771)
        (3.0,22.9851741)
        (4.0,36.7058897)

    };
    \addlegendentry{Relative RMSE}
    \end{axis}
\end{tikzpicture}
    \caption{Relative MAE et RMSE scores for the Random Forest's predictions compared to the base bitrate.}
    \label{fig:relative}
\end{figure}
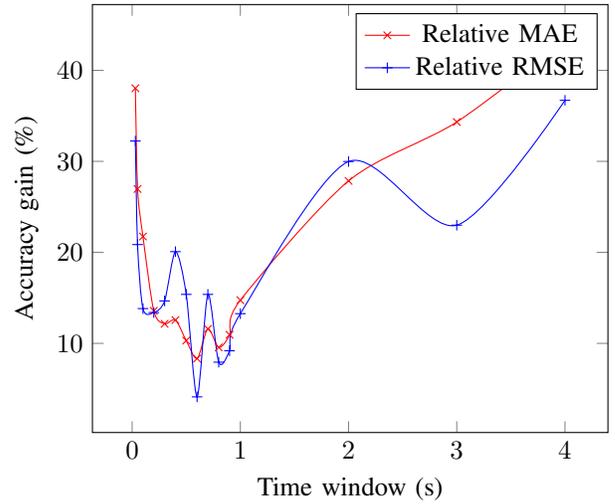

Another interesting observation can be made from Fig. \ref{fig:relative}: the accuracy gained by the Random Forest model compared to simply using the current bitrate (i.e., no forecasting) is higher when the time windows are $\leq$ 0.1s and $\geq$ 2.0s. These experimental results show that this framework is pertinent, with an average accuracy gain of 19.73\% in MAE and 18.00\% in RMSE.

% \begin{figure}[!htb]
%     \centering
%     \input{figures/engineering4}
%     \caption{Accuracy gain with and without feature engineering.}
%     \label{fig:engineering}
% \end{figure}

%%%%%%%%%%%%%%%%%%%%%%%%%%%%%%%%%%%%%%%%%%%%%%%%%%%%%%%%%%%%%%%%%%%%%%%%%%%%%%%%%%%%%%%%%%%%%%%%%%%%%%%%%%

\section{Conclusion}\label{sec:conclusion}
In this paper, a framework to predict future traffic flows' bandwidth has been proposed. The feature extraction and engineering process from raw network packets to a flow format specifically designed for this task has been detailed and evaluated. Four machine learning models (DT, RF, XGBoost, DNN) have been retained, according to two metrics (MAE and RMSE) and we conclude that the Random Forest is the most efficient algorithm for bandwidth forecasting. The framework provides two promising results: i) predictions are 19.73\% more accurate (according to MAE) on average compared to no forecasting and ii) through all our experiences, RF's predictions are never worse than no forecasting.

Despite this good performance, the framework only handles ongoing flows and flows that end during the next time window. New flows that appear during the next time window are not predicted yet by the framework, thus this work can be considered as a work-in-progress. As for future work, we envision testing other machine learning models (like Generative Adversarial Networks) to learn the behavior of flows creation to predict realistic new flows at each time window.

%%%%%%%%%%%%%%%%%%%%%%%%%%%%%%%%%%%%%%%%%%%%%%%%%%%%%%%%%%%%%%%%%%%%%%%%%%%%%%%%%%%%%%%%%%%%%%%%%%%%%%%%%%

% \bibliographystyle{IEEETran}
% \bibliography{references}


\end{document}